\documentclass[twocolumn,superscriptaddress,amsmath,amssymb,showpacs,showkeys]{revtex4}

\usepackage{graphicx,color}
\usepackage{dcolumn}
\usepackage{bm}
\usepackage{enumerate} 
\newcommand{\beq}{\begin{equation}} \newcommand{\eeq}{\end{equation}}
\newcommand{\bqa}{\begin{eqnarray}} \newcommand{\eqa}{\end{eqnarray}}

\definecolor{gold}{rgb}{0.75,0.56,0.00}
\definecolor{green}{rgb}{0.00,0.50,0.00}

\newcommand{\ms}[1]{\mbox{\scriptsize #1}}

\begin{document}


\title{Multi-parameter estimation along quantum trajectories with Sequential Monte Carlo methods}

\author{Jason F. Ralph}
 \email{jfralph@liverpool.ac.uk}
 \affiliation{Department of Electrical Engineering and Electronics, University of Liverpool,  Brownlow Hill, Liverpool, L69 3GJ, UK.}
\author{Simon Maskell}
 \email{smaskell@liverpool.ac.uk}
 \affiliation{Department of Electrical Engineering and Electronics, University of Liverpool,  Brownlow Hill, Liverpool, L69 3GJ, UK.}
\author{Kurt Jacobs}
 \email{kurt.jacobs@umb.edu}
  \affiliation{U.S. Army Research Laboratory, Computational and Information Sciences Directorate, Adelphi, Maryland 20783, USA.}
  \affiliation{Department of Physics, University of Massachusetts at Boston, Boston, MA 02125, USA} 
\affiliation{Hearne Institute for Theoretical Physics, Louisiana State University, Baton Rouge, LA 70803, USA} 

\date{\today}

\begin{abstract}
This paper proposes an efficient method for the simultaneous estimation of the state of a quantum system and the classical parameters that govern its evolution. This hybrid approach benefits from efficient numerical methods for the integration of stochastic master equations for the quantum system, and efficient parameter estimation methods from classical signal processing. The classical techniques use Sequential Monte Carlo (SMC) methods, which aim to optimize the selection of points within the parameter space, conditioned by the measurement data obtained. We illustrate these methods using a specific example, an SMC sampler applied to a nonlinear system, the Duffing oscillator, where the evolution of the quantum state of the oscillator and three Hamiltonian parameters are estimated simultaneously. 
\end{abstract}


\maketitle

\section{Introduction}\noindent
Stochastic master equations provide a model for the evolution of open quantum systems subject to continuous measurements~\cite{Bel1999,Wis2010,Jac2014}. The trajectories that the stochastic master equations generate represent the evolution of the state of an individual quantum system, conditioned on a particular measurement record. In theoretical studies, the measurement record is a simulated sequence corresponding to a particular realization of the evolution. However, recent experiments that implement continuous quantum measurements have demonstrated that the evolution of individual quantum systems can be reconstructed from experimental data~\cite{Mur2013,Web2014,Six2015,Cam2016}. The generation of such trajectories in real time during the measurement process will be an important step towards state-dependent feedback control of individual quantum systems~\cite{Bel1999,Wis2010,Jac2014}. Feedback control has been demonstrated in quantum systems using the output measurement record as an input signal to the control system in optical~\cite{Smi2002, Bra2012}, opto-mechanical~\cite{Kub2009, Wil2015, Sud2017}, and mesoscopic superconducting systems~\cite{Vij2012, Ris2012}. In most of these examples, while the direct use of the measurement record in the control system demonstrates the utility of quantum feedback control, it is limited by the fact that the evolution of the underlying state of the system is not included in the generation of the controls. State-dependent control is more flexible and can include quantities that are estimated from, but are not directly measured in experiments. 

Given a measurement record, a Stochastic Master Equation (SME) provides an estimate of the quantum state at each point in time, and -- in the case of mixed states -- an indication of the uncertainty associated with this state in terms of an estimate of its purity. The SME is derived by taking a single quantum system and coupling it weakly to environmental degrees of freedom that mediate a continuous measurement process. The continuous measurement of the coupled system is realized by continuously measuring the state of the environment, which can be modeled as a sequence of projective measurements on successive environmental degrees of freedom. The simplest example of this process is the measurement of the electromagnetic radiation emitted by the system, in which case the environment is the electromagnetic field~\cite{Wis2010,Jac2014}. In the most common form of the SME, a Markovian condition is applied, meaning that the environment carries information away from the system but does not by itself feed this information back to affect the system at a later time. The resulting evolution of the system is continuous and stochastic, with the stochastic term arising from the effect of the sequence of measurements on the combined system. 

In simulations, an SME is used to analyze the properties of an open system under the action of a continuous sequence of measurement operators, using a realization for the noise process and calculating the evolution of the system conditioned on this realization. When interpreting experiments, the SME is used to reconstruct the estimate of the quantum state as a function of time from the given measurement record provided by the experiment. In this regard, the SME is very similar to classical state estimation techniques (often referred to as `target tracking' or `object tracking' \cite{Bla1986,Bar2001,Ral2010}), which are used to interpret sequences of classical noisy sensor measurements to form a coherent picture of the world. These classical techniques have been developed to interpret sensor data (radar or sonar signals, and sequences of images) where objects are moving against noisy backgrounds. The motion of the objects may be unpredictable or uncooperative, their identity may not be known from the measurements, they may be occluded for periods of time, and individual objects may not be fully resolved by the sensor. Classical state estimation techniques provide methods to solve all of these problems and ambiguities.

With a continuous quantum measurement, the measurement record contains information about the evolution of the particular quantum state (a {\em quantum trajectory}) but the properties of this trajectory also contain information about the classical parameters that govern the dynamics of the system: the classical parameters in the Hamiltonian and the strength of the coupling to the environment. In this paper, we demonstrate how the stochastic master equation can be augmented with techniques drawn from classical state estimation, Sequential Monte Carlo (SMC) methods, to estimate several Hamiltonian parameters efficiently alongside the quantum trajectories. 

We begin our presentation by first reviewing other approaches to Hamiltonian parameter estimation, and the development of a set of Hybrid SMEs for the quantum evolution and the Kushner-Stratonovich equation for the classical parameters \cite{Ral2011} in sections II and III, respectively. In section IV, we then discuss how efficient classical parameter estimation techniques \cite{Gre2017} can be applied to the solution of the classical aspects of the Hybrid SME. Section V introduces an example system, the Duffing oscillator, which contains a number of relevant experimental parameters, and Section VI presents results for the simultaneous estimation of the quantum trajectories for the oscillator and up to three Hamiltonian parameters. Section VII discusses how such methods may be useful in practical systems and draws conclusions from the results presented. 

\section{Hamiltonian Parameter Estimation}\noindent

The problem of estimating the dynamical parameters of a quantum system has been studied previously by a number of authors, including those who have adopted a continuous measurement approach. This estimation process is often referred to as Hamiltonian parameter estimation. This is because the basic description of quantum dynamics is encapsulated by the Hamiltonian operator, which determines the equations of motion, and values of the classical parameters in the Hamiltonian determine the specifics of the evolution. 

A standard method for determining the dynamics of a quantum system is to prepare it many times in a range of different initial states, allow it to evolve, and then measure it before re-preparation. The results of the measurements can then be combined in a tomographic-like process to obtain the equation of motion for linear Schr\"{o}dinger evolution~\cite{Chu1997}. An alternative approach, and the one in which we are interested here, is to prepare the system only once and to continually monitor its subsequent evolution to build a picture of its dynamics. A full description of the problem involves starting with a prior probability density for the parameters one wishes to determine and then using Bayes' theorem to continually update this probability density from the stream of measurement results as they are obtained. A number of authors have considered this problem~\cite{Gam2001, Ver2001, Sto2004, Tsa2009a, Tsa2009b, Tsa2010, Tsa2011, Ral2011, Neg2013, Ber2015, Bas2015, Cor2017}. This is of particular interest when the parameters of a system change slowly with time, and one wishes to be able to track the variations in the parameters. It is also relevant to the problem of using quantum systems as probes to measure time-varying classical fields (such as gravity waves~\cite{LIGO2016} and magnetic fields~\cite{Yan2017}), as these fields appear as parameters in the Hamiltonian. 

As discussed in the introduction, a dynamical equation referred to as the stochastic master equation (SME) can be used to track the evolution of a quantum system from the results of a continuous measurement so long as the dynamical parameters of the system are known. If they are not known then the full estimation problem involves both the SME and a Kushner-Stratonovich equation that evolves the probability density for the parameters of the system. The combined set of dynamical equations has been referred to as a {\em Hybrid SME}~\cite{Ral2011}. The first papers on the subject of Hamiltonian parameter estimation via continuous measurements were concerned mainly with deriving the Hybrid SME and applying it to the estimation of a single parameter~\cite{Gam2001, Ver2001}. Subsequently, Tsang and collaborators considered the more general problem of smoothing in which a time-varying parameter (a signal or wave-form) is estimated from all the measurement results obtained, and determined the ultimate limits to this procedure~\cite{Tsa2009a, Tsa2009b, Tsa2010, Tsa2011, Ber2015}. An alternative and interesting approach to the problem was proposed recently by Bassa~\textit{et al.\ }\cite{Bas2015}. While most of the related work on parameter estimation employs continuous measurements, this approach considered a sequence of instantaneous measurements, and employed a discrete version of the Hybrid SME where several classical parameter values were encoded in an expanded quantum state. 

A major problem with the Hybrid SME is that it is highly demanding from a computational point of view; in order to evolve the Kushner-Stratonovich equation for the probability density describing the observer's knowledge of the parameters, the SME must be evolved for every value of the parameters for which this density is appreciable. The grid of points for which the SME must be evolved becomes large very quickly as the number of parameters increases. Two previous papers have put forward methods aimed at addressing this difficulty. Ralph~\textit{et al.\ }\cite{Ral2011} and Cortez~\textit{et al.\ }\cite{Cor2017} considered the estimation of a single frequency parameter, and presented methods to bypass the Kushner-Stratonovich equation. These papers estimate the natural oscillation frequency of a qubit directly from the measurement record. This approach has many benefits in terms of computational efficiency, but it has the disadvantage of not providing a simultaneous estimate of the quantum state of the system -- which would be provided by the full solution of the Hybrid SME. Here, we will explore the use of a potentially more powerful technique in which the probability density is replaced with a finite set of samples of the parameters that are evolved instead. The examples given below typically use 50-100 quantum states and the equivalent of thousands to millions of classical parameter values. The purpose is again to reduce the number of copies of the quantum state that must be evolved in parallel using the SME, but we will apply this method to the challenging problem of estimating multiple parameters simultaneously.

\section{Hybrid Stochastic Master Equations}\noindent
The simultaneous estimation of the quantum state of a system and the classical Hamiltonian parameters that govern its evolution was considered in Ref.~\cite{Ral2011}, where an approach was presented based on a set of parallel SMEs, each using a different set of parameter values contained in a vector $\underline{\lambda}$, which have an associated probability. The final mixed state is then constructed by averaging over the probabilities for the classical parameters. The probabilities associated with the different parameter vectors evolve via a Kushner-Stratonovich equation and are conditioned on the continuous measurement record \cite{Ral2011}. 

For a quantum system subject to a continuous measurement, with a known set of Hamiltonian parameters, the evolution of the quantum state, $\rho_c(t)$, conditioned on the measurement record, $y(t)$, is given by the stochastic master equation~\cite{Bel1999,Wis2010,Jac2014}. In general, the interaction with the environment can be represented by a set of system operators which are coupled to environmental degrees of freedom, some of which are not measured $\hat{V}_j$  $(j = 1\dots m)$ (`unprobed' operators), and some of which are measured and generate the continuous weak measurement $\hat{L}_r$ $(r = 1\dots m')$. In an ideal case, the measurement record is 100\% efficient, with all of the available information being reflected in the measurement record. Unfortunately, real measurements are rarely ideal and the continuous measurement record is often corrupted with extraneous (classical) noise sources. These extraneous degrees of freedom can be characterized by an efficiency parameter for the measurement operators, $\hat{L}_r$ has an efficiency $\eta_r$. Specifically, $\eta_r$ is the fraction of the total noise power due to the quantum measurement as opposed to power contained in the other extraneous noise sources. 

For unprobed operators $\hat{V}_r$ and measurement operators $\hat{L}_r$, the general form for the SME is given by,
\begin{eqnarray}\label{sme1}
d\rho_c&=&- i \left[\hat{H},\rho_c\right]dt \nonumber\\
&&+\sum_{j=1}^{m} \left\{ \hat{V}_{j} \rho_c \hat{V}^{\dagger}_{j} -\frac{1}{2}\left(\hat{V}^{\dagger}_{j} \hat{V}_{j} \rho_c 
+ \rho_c \hat{V}^{\dagger}_{j} \hat{V}_{j} \right)\right\}dt  \nonumber\\
&&+\sum_{r=1}^{m'} \left\{ \hat{L}_{r} \rho_c \hat{L}^{\dagger}_{r} -\frac{1}{2}\left(\hat{L}^{\dagger}_{r} \hat{L}_{r} \rho_c 
+ \rho_c \hat{L}^{\dagger}_{r} \hat{L}_{r} \right)\right\}dt   \nonumber\\
&&+ \sum_{r=1}^{m'} \sqrt{\eta_r}\left(\hat{L}_{r}\rho_c+\rho_c \hat{L}^{\dagger}_{r}-\mathrm{Tr}(\hat{L}_{r}\rho_c+\rho_c \hat{L}^{\dagger}_{r}) \right)dW_{r} \nonumber\\
\end{eqnarray}
where $\hat{H}$ is the Hamiltonian of the system, $dt$ is an infinitesimal time increment, and the measurement record for each of the measurement operators $\hat{L}_{r}$ during a time step $t\rightarrow t+dt$ is given by, $y(t+dt)-y(t)=dy_r(t)= \sqrt{\eta_j}\mathrm{Tr}(\hat{L}_{r}\rho_c+\rho_c \hat{L}^{\dagger}_{r}) dt+dW_{r}$. We will take $dW_{r}$ to be a real Wiener increment such that $dW_{r}=0$ and $dW_{r} dW_{r'}  = \delta_{rr'}dt$ for simplicity, but this is not strictly necessary. More general forms of complex increments may also be used~\cite{Wis2005}. 

Where the evolution of a quantum system is governed by a set of Hamiltonian parameters that are not known exactly, we can describe the parameters in terms of a classical probability density, $P(\underline{\lambda})$, where $\underline{\lambda}=(\lambda_1, \lambda_2,...)$. The system is then described by a set of SMEs, one for each set of possible parameter values,
\begin{eqnarray}\label{sme1a}
d\rho_{c,\underline{\lambda}}&=&- i \left[\hat{H}(\underline{\lambda}),\rho_{c,\underline{\lambda}}\right]dt \nonumber\\
&&+\sum_{j=1}^{m} \left\{ \hat{V}_{j} \rho_{c,\underline{\lambda}} \hat{V}^{\dagger}_{j} -\frac{1}{2}\left(\hat{V}^{\dagger}_{j} \hat{V}_{j} \rho_{c,\underline{\lambda}} 
+ \rho_{c,\underline{\lambda}} \hat{V}^{\dagger}_{j} \hat{V}_{j} \right)\right\}dt  \nonumber\\
&&+\sum_{r=1}^{m'} \left\{ \hat{L}_{r} \rho_{c,\underline{\lambda}} \hat{L}^{\dagger}_{r} -\frac{1}{2}\left(\hat{L}^{\dagger}_{r} \hat{L}_{r} \rho_{c,\underline{\lambda}} 
+ \rho_{c,\underline{\lambda}} \hat{L}^{\dagger}_{r} \hat{L}_{r} \right)\right\}dt   \nonumber\\
&&+ \sum_{r=1}^{m'} 
\sqrt{\eta_r}\left(\begin{array}{l}
\hat{L}_{r}\rho_{c,\underline{\lambda}}+\rho_{c,\underline{\lambda}} \hat{L}^{\dagger}_{r}  \\
 -\mathrm{Tr}(\hat{L}_{r}\rho_{c,\underline{\lambda}}+\rho_{c,\underline{\lambda}} \hat{L}^{\dagger}_{r}) dW_{r}
\end{array}\right)
\end{eqnarray}

The evolution of the probability density $P(\underline{\lambda})$ is governed by a Kushner-Stratonovich stochastic differential equation, derived in \cite{Ral2011} for a single efficient measurement operator and in the absence of additional unprobed environmental operators. Any unprobed environmental operators affect the evolution of the individual SMEs but they do not play a role in the evolution of $P(\underline{\lambda})$. However, the equation given in \cite{Ral2011} generalizes naturally to include measurement inefficiencies and is given by,
\begin{eqnarray}\label{KS1}
dP_r(\underline{\lambda}) = &\sqrt{\eta_r}(\mathrm{Tr}(\hat{L}_r\rho_{c,\underline{\lambda}}+\rho_{c,\underline{\lambda}}\hat{L}^{\dagger}_r)-\mathrm{Tr}(\hat{L}_r\rho_c+\rho_c\hat{L}^{\dagger}_r)) \nonumber \\
&\times(dy_r(t)-\sqrt{\eta_r}\mathrm{Tr}(\hat{L}_r\rho_c+\rho_c\hat{L}^{\dagger}_r) dt)P(\underline{\lambda}) 
\end{eqnarray} 
where $dP_r(\underline{\lambda})$ is the update to the probability density due to a measurement increment $dy_r(t)$ corresponding to the measured operator $\hat{L}_{r} $, such that
\begin{equation}\label{dP}
P(dy_r(t)|\underline{\lambda})=\frac{e^{(-(dy_r(t)-\sqrt{\eta}\mathrm{Tr}(\hat{L}_r\rho_{c,\underline{\lambda}}+\rho_{c,\underline{\lambda}}\hat{L}^{\dagger}_r))^2/dt)}}{\sqrt{2\pi dt}}
\end{equation} 
and the full conditional density matrix $\rho_c$ is given by,
\begin{equation}\label{full_rho}
\rho_c = \int_{\underline{\lambda}}P(\underline{\lambda})\rho_{c,\underline{\lambda}} d{\underline{\lambda}}
\end{equation} 

\section{Sequential Monte Carlo Methods}\noindent
Sequential Monte Carlo methods originate in the field of multi-target tracking~\cite{Gor1993}), but have been adopted and generalized to form a set of very efficient methods for parameter and state estimation in classical signal processing and nonlinear filtering. SMC methods are sometimes referred to as {\em particle filters}, but particle filters are a special case of the general approach. An SMC method relies on the approximation of a continuous probability distribution by a finite set of points (or particles) which sample the parameter space. The importance of each sample point changes in response to (is conditioned by) the measurements associated with the parameters being estimated, and the sample points can be periodically resampled to concentrate sampling towards regions of higher relative probability. 

In particle filters, the sample points are allowed to evolve according to some dynamical process, generating a time dependent history or a track within the parameter space. In the example presented in this paper, the parameters are selected to be constant and another SMC method is more suitable. We adopt an approach used recently for parameter estimation in classical differential equations \cite{Gre2017}, which is an example of a {\em Sequential Monte Carlo sampler}. This approach is particularly well-suited to the estimation of fixed parameters; however, the SMC sampler used here still embodies all of the key features of a general SMC method: sampling, conditioning/updating, and resampling. A number of very approachable tutorials and introductions to particle filters and general SMC methods have been published. For example, a comprehensive guide to SMC methods and their applications is available in \cite{Dou2001}, a mathematical introduction is given in \cite{Cap2007}, and a widely cited tutorial to particle filters and SMC methods is contained in \cite{Aru2002}.

Formally, an SMC method approximates a (classical) expectation for a function $h(x)$ over a probability distribution $p(x)$ defined on some parameter space $\Lambda$, $x \in \Lambda$, given by
$$
\bar{h} = \int h(x)p(x)dx
$$
using a finite sum of a set of points $x^{(i)}$ ($i = 1\ldots N$) drawn from $p(x)$, which is known as the {\em target distribution}. The expectation value for an arbitrary function can be approximated by, 
$$
\bar{h} \simeq \frac{1}{N}\sum_{i =1}^{N}h(x^{(i)})
$$
The larger the number of sample points, the better the approximation -- in fact, under reasonable assumptions, the variance of the error in $\bar{h}$ can be shown to scale as 1/N in any number of dimensions \cite{Pre2007}. The problem is that, in most practical cases, the probability distribution is unknown. It needs to be estimated from a sequence of measurements. To do this, another distribution, the {\em proposal distribution} $q(x)$, is introduced such that \cite{Cap2007},
\begin{eqnarray}
\bar{h}& = &\int h(x)q(x)\frac{p(x)}{q(x)}dx=\int h(x)q(x)w(x)dx \nonumber \\
&&\simeq\sum_{i =1}^{N}\frac{w^{(i)}}{\sum_{j =1}^{N}w^{(j)}}h(x^{(i)}) 
\end{eqnarray}
where $w(x)=p(x)/q(x)$ and the $w^{(i)}$'s are (unnormalized) weights associated with each sample point. In our case, each sample point is associated with a parameter value or a vector of values for each of the parameters being sought, $w^{(i)} \leftrightarrow \underline{\lambda}^{(i)}$, where $\underline{\lambda}^{(i)} \in \Lambda$. Initially, the sample points are randomly selected from a prior distribution, which covers the entire range of possible parameter values, and are given a uniform weight. As new measurements are added, the accuracy of the estimated quantity $\bar{h}$ is improved by updating the weights associated with the particles to reflect the new information that the measurement contains. Some weights are increased and some weights are reduced when the measurement supports or contradicts the corresponding sample point, respectively. 

The values $w^{(i)}$ are referred to as `weights' rather than probabilities because, although they are related to probabilities, they are not necessarily normalized after each time step and do not necessarily sum to one. In practice, it is convenient to normalize the weights after each time step. Here, we denote the normalized weights by $\tilde{w}^{(i)}$ \cite{Gre2017}. When sample points have very low weight, and hence very low probability, they can be removed and replaced with alternative particles, but this resampling process must be done carefully so as to ensure that the statistical quantities remain unbiased and will converge efficiently to the desired values. 

The proposal distribution should be simple to calculate and different choices of $q(x)$ are used in different variants of the SMC approach \cite{Dou2001,Cap2007,Aru2002}. The secret to working with SMC methods is to pick a suitable proposal distribution to solve the problem in a robust manner using limited computational resources. In particular, a good choice of proposal distribution allows classical parameters to be estimated significantly more efficiently than when using an enumerative or grid based method \cite{Aru2002}, as was used for a one parameter Hybrid SME problem in \cite{Ral2011,Bas2015}. 

For our Hybrid SME problem, we start by selecting an initial set of sample points in the parameter space using a prior distribution and initialize an SME (\ref{sme1a}) for each of the sample points. For the examples shown below, the quantum state of the system is initialized to be a thermal mixed state, and the prior distribution is chosen to be uniform over some finite range within which the true parameter values are known to lie. An accurate initial prior distribution can significantly reduce the number of particles required by the SMC sampler, but in many situations the prior is not well defined. Once the points have been selected, the weights are initialized with $w^{(i)}_0 = \tilde{w}^{(i)}_0 = 1/N$.  

For each time step, the individual SMEs are integrated using the increment (\ref{sme1a}) found using the parameter value $\underline{\lambda}^{(i)}$ associated with the particle. A corresponding measurement probability is found from (\ref{dP}) and used to update the (unnormalized) particle weight $w^{(i)}_{k-1,r} \rightarrow w^{(i)}_{k,r}$ using \cite{Gre2017}
\begin{equation}\label{weights} 
w^{(i)}_{k,r}=p(dy_r(t_k)|\underline{\lambda}^{(i)})w^{(i)}_{k-1,r}
\end{equation}
for the $k$'th measurement from measurement operator $\hat{L}_r$ at time $t_k$. All particles are updated after the measurement increment and then weights are normalized. 

Resampling to generate new particles only occurs when the distribution of weights amongst the particles is such that the effective sample size (or the effective number of particles) $N_{eff} = 1/(\sum_i (\tilde{w}^{(i)})^2)$ falls below some threshold value -- indicating that the weight is being concentrated in a relatively small number of particles and a significant number of particles have low weight and do not contribute to the estimates; a problem known in the SMC literature as {\em sample impoverishment} or {\em weight degeneracy} \cite{Cap2007}. It is known that the variance of the weight distributions across different realizations of the SMC sampler is guaranteed to grow with each time step \cite{Aru2002}. However, since a given realization (i.e. one run of the algorithm) does not have access to the ensemble of all possible realizations, it is convenient to monitor something that can be computed from a single realization. The effective sample size is well established as such a quantity \cite{Cri2002} and it can be considered to be a noisy measurement of the (inverse of the) variance. Between resampling events, the variance of the weight distributions will (on average) increase and the effective sample size will decrease. While the precise threshold value used is a somewhat arbitrary choice for the algorithm designer, it is common (across the vast range of applications of SMC samplers and particle filters) to consider threshold values between $N/10$ and $N/2$. In the cases shown below the threshold value for $N_{eff}$ was set to be $N/2$ \cite{Gre2017}. 

When the particles are resampled, the new candidate values $\tilde{\underline{\lambda}}$ are sampled from the distribution formed from the current particle weights. The particles with the highest weights are more likely to be selected, although the particles with relatively low weights still have a chance of being selected. The new particle parameter values are then selected using the distribution $q(\tilde{\underline{\lambda}}|\underline{\lambda}^{(i)})={\cal N}(\tilde{\underline{\lambda}};\underline{\lambda}^{(i)},\Sigma)$, where ${\cal N}(x;\mu_{x},S)$ is a normal distribution with mean $\mu_{x}$ and covariance $S$, and $\Sigma$ is related to the covariance matrix associated with the current particle weights, $\Sigma_k$. The role of $q(\tilde{\underline{\lambda}}|\underline{\lambda}^{(i)})$ is to select new points, $\tilde{\underline{\lambda}}^{(i)}$, around the current particles with large weights, but not at exactly the same point. In this paper, we use a {\em defensive strategy} \cite{Gre2017,Hes1995}, where 90\% of resampled points use a covariance which is 10\% of the current covariance, $\Sigma=0.1\Sigma_k$, and 10\% of the resampled points use the full covariance matrix $\Sigma=\Sigma_k$. This allows for small perturbations in parameter space around the high weight sample points, including the correlations between different parameters seen in the covariance matrix, and a small number of large excursions, to explore more of the parameter space than is currently being covered by the sample points with large weight. There are two specific design considerations relevant to the choice of the distribution, $q$. The first is to ensure that having more samples will give rise to more accurate estimates of quantities of interest, which is manifest empirically as robustness. Put simply, this demands that samples are proposed in a way that explores possible but potentially low probability states. The second is to ensure that the SMC method is computationally efficient, i.e. that it gets as accurate an estimate as is possible with a given number of samples. This demands that samples are placed in high probability areas. A defensive proposal is an advanced, but relatively standard technique, used in particle filters and SMC samplers, that combines robustness with efficiency by having two elements to the proposal, one that is designed to ensure the sampler is robust and the other that is designed to ensure that it is efficient.

When the new sample points have been selected, they are initially assigned the weight $\check{w}^{(i)}_0 = 1/N$ and then the unnormalized weight for the new candidate points is calculated reusing the entire record of measurement increments,
\begin{equation}\label{resample}
\check{w}^{(i)}_k=\frac{\prod_{r,k}p(dy_r(t_k)|\tilde{\underline{\lambda}}^{(i)})}{\prod_{r,k}p(dy_r(t_k)|\underline{\lambda}^{(i)})}\check{w}^{(i)}_0
\end{equation}
Once all of the new weights have been recalculated they can be renormalized, and the integration of the Hybrid SME can continue as normal. Where the new sample weights are still degenerate and the effective number of particles is still below the threshold value, the resampling (and recalculation) needs to be performed again. 
\begin{figure}[htbp]\label{process}
	\centering
		\includegraphics[width=1.0\hsize]{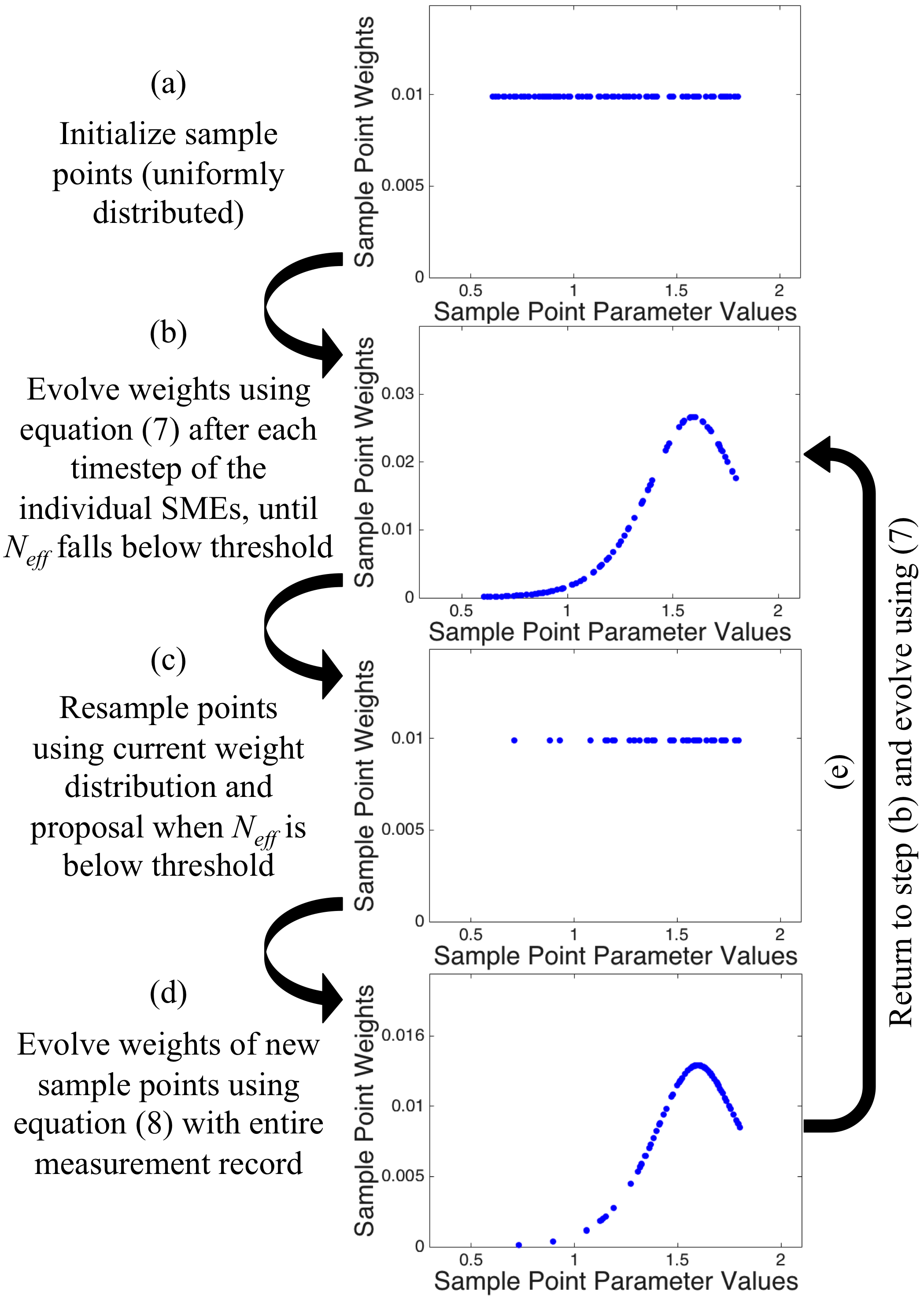}
	\caption{\label{fig:SMCSteps} (Color online) Schematic process, showing the main steps in the SMC Sampler for a one parameter example.}
\end{figure}

To summarize, the SMC method used here can be described in the following five steps:
\begin{enumerate}[(a)]
\item Initialize the individual density matrices $\rho_{c,\underline{\lambda}^{(i)}}$ using thermal mixed states, and select a set of classical parameter values (sample points or particles) $\underline{\lambda}^{(i)}$ using a uniform distribution covering the full range of possible parameter values, and assign uniform weights to each of these $\tilde{w}^{(i)}_0 = 1/N$.
\item Evolve the quantum state using the individual SMEs (\ref{sme1a}), using the corresponding classical parameters and updating their weights using (\ref{weights}). Continue this evolution until $N_{eff}$ drops below the threshold value.
\item If $N_{eff}$ is below the threshold value, the classical parameter values are resampled using a cumulative probability distribution calculated from the particle weights. This resampling creates a new set of particles/sample points, where the classical parameters are selected around the `parent' values. The defensive strategy introduces small perturbations around the parent values and the occasional large perturbation to explore a wider parameter space -- the new weights associated with each of the new parameter values/sample points are uniformly distributed at this point.
\item Once the new values have been selected, the complete evolution of each quantum state is recalculated using new initial thermal states and the individual SMEs (using the same measurement record), and the uniform weights from step (c) are recalculated using (\ref{resample}). 
\item Return to step (b) with evolution of the quantum state and weights determined by the individual SMEs and the weight update (\ref{weights}), until $N_{eff}$ drops below the threshold value again, at which point the resampling step (c) and the re-weighting step (d) are again required. 
\end{enumerate}
A schematic example of the estimation process for a one dimensional parameter example is shown in Figure 1. In this example, it is possible to see that the initial uniform weighting of the particles evolves so that the relatively large number of particles below a parameter value of $1.0$ carry very little weight, and the distribution of particles immediately after resampling is concentrated more towards the values above 1.0. The re-weighted parameter values shown in (d) represent a better approximation to the underlying probability distribution than those shown in (b), which contains significant gaps towards the peak of the distribution. For a more detailed description of the implementation, a full description of the SMC method is given as pseudo-code in \cite{Gre2017}.

The recalculation over the entire measurement history is an unfortunate, but necessary, computational cost in the SMC sampler. Recalculating the weights for the entire history of measurement increments will often take a significant amount of time. However, the need to regularly resample the entire set of particles reduces as the distribution of the particles improves to reflect the information contained in the measurements \cite{Gre2017}. This means that the computational load introduced is biased towards the start of the calculation of a quantum trajectory. In addition, for resampled points very close to the parent particles, some approximations are possible based on the fact that the ratio between the products in (\ref{resample}) is very close to one. It is not possible to remove the recalculation entirely however without constraining the resampled parameter values and therefore not exploring the full parameter space. 

\section{Example System -- Duffing Oscillator}\noindent
The properties of the quantum trajectories generated by the Duffing oscillator have been studied extensively in terms of the appearance of chaotic behavior from quantum systems in the classical limit \cite{Sch1995,Bru1996,Bru1997,Hab1998,Bha2000,Sco2001,Bha2003,Eve2005,Eas2016,Pok2016,Ral2017}, but it is also a model used for a number of other practical systems where quantum effects in classical nonlinear systems are of interest. For example, it has been used to describe the motion of a levitated particle in an electromagnetic trap \cite{Gie2013,Gie2015}, and is the basis for the analysis of the properties of vibrating beam accelerometers \cite{Aik2001,Mes2009,Agr2013}. The Hamiltonian for the Duffing oscillator can be written in the general form, using dimensionless position and momentum operators $\hat{q}$ and $\hat{p}$,
\begin{equation}\label{DuffHam}
\hat{H}(\underline{\lambda}) = \frac{1}{2}\hat{p}^2+\frac{1}{2}\omega^2\hat{q}^2+\frac{1}{4}\mu\hat{q}^4+g\cos(t)\hat{q}+\frac{\Gamma}{2}(\hat{q}\hat{p}+\hat{p}\hat{q})
\end{equation}
where the vector $\underline{\lambda}=(\omega,\mu,g)$ contains the three Hamiltonian parameters of interest: the natural (linear) oscillation frequency $\omega$, the nonlinear coefficient $\mu$, and the strength of the external driving term $g$. The measurement is applied via a Linblad operator $\hat{L}=\sqrt{2\Gamma}\hat{a}$, and $\hat{a}$ is the harmonic oscillator lowering operator so that $\hat{q}=(\hat{a}^{\dagger}+\hat{a})/\sqrt{2}$ and $\hat{p}=i(\hat{a}^{\dagger}-\hat{a})/\sqrt{2}$ with $[\hat{a},\hat{a}^{\dagger}]=1$ and $\hbar=1$. We fix the measurement strength so that $\Gamma=0.125$ for all of the results presented here. The final term in the Hamiltonian is included because, in combination with the dissipative measurement process, it generates linear damping in momentum. This is a useful numerical addition because it keeps the phase space contained, thereby restricting the numbers of states required in the simulation, without affecting the underlying physics. 

The numerical integration of the individual SMEs uses a method developed by Rouchon and colleagues \cite{Ami2011,Rou2015} specifically for stochastic master equations. This method has been demonstrated to provide significant benefits in terms of accuracy versus computational resources when compared to standard methods, such as Milstein's method \cite{Mil1995}, for both systems involving small numbers of basis states \cite{Rou2015} and large numbers of basis states \cite{Ral2016}. We also employ a moving basis method used by Schack, Brun and Percival \cite{Sch1995,Bru1996} to shift basis states to be centred on the current expectation value of the state. Although not strictly necessary \cite{Sch1995,Bru1996}, we shift the basis after each time step. This comes at a computational cost but it also ensures that the number of basis states employed is minimized. Once the evolution of the individual SMEs has been calculated, using the appropriate set of parameters, the combined density operator is calculated by averaging over all of the individual states, weighted appropriately by the particle weights. 

The increment to the state $\rho_{\ms{c},\underline{\lambda}}^{(n)}$ for the time step from $t_n = n\Delta t$ to $t_{n+1} = (n+1)\Delta t$ is calculated using
\begin{equation}\label{sme2}
\rho_{\ms{c},\underline{\lambda}}^{(n+1)}= \frac{\hat{M}_{n,\underline{\lambda}}\rho_{\ms{c},\underline{\lambda}}^{(n)}\hat{M}_{n,\underline{\lambda}}^{\dagger}+(1-\eta)\hat{L}\rho_{\ms{c},\underline{\lambda}}^{(n)}\hat{L}^{\dagger}\Delta t }
{\mathrm{Tr}\left[\hat{M}_{n,\underline{\lambda}}\rho_{\ms{c},\underline{\lambda}}^{(n)}\hat{M}_{n,\underline{\lambda}}^{\dagger}+ (1-\eta)\hat{L}\rho_{\ms{c},\underline{\lambda}}^{(n)}\hat{L}^{\dagger}\Delta t \right]}
\end{equation}
where $\Delta \rho_{\ms{c},\underline{\lambda}}^{(n)} = \rho_{\ms{c},\underline{\lambda}}^{(n+1)}- \rho_{\ms{c},\underline{\lambda}}^{(n)}$ and $\hat{M}_{n,\underline{\lambda}}$ is given by 
\begin{eqnarray}\label{Mn1}
\hat{M}_{n,\underline{\lambda}} &=& I-\left(i\hat{H} +\frac{1}{2} \hat{L}^{\dagger}\hat{L}\right)\Delta t +\frac{\eta}{2}\hat{L}^2(\Delta W(n)^2-\Delta t)  \nonumber\\
&& +\sqrt{\eta}\hat{L}\left(\sqrt{\eta}\mathrm{Tr}[\hat{L}\rho_{\ms{c},\underline{\lambda}}^{(n)}+\rho_{\ms{c},\underline{\lambda}}^{(n)}\hat{L}^{\dagger}]\Delta t +\Delta W(n)\right) \nonumber
\end{eqnarray}
where the $\Delta W$'s are independent Gaussian variables with zero mean and a variance equal to $\Delta t$. Once the increment has been calcluated, center of the basis is moved to the new location of the state in phase space, as given by the expectation values of the phase space operators, $(q_{(n+1),\underline{\lambda}}, p_{(n+1),\underline{\lambda}}) = (\mathrm{Tr}[\hat{q}\rho_{\ms{c},\underline{\lambda}}^{(n+1)}], \mathrm{Tr}[\hat{p}\rho_{\ms{c},\underline{\lambda}}^{(n+1)}])$, using the displacement operator \cite{Sch1995,Bru1996},
\begin{equation}
\hat{D}(p_{(n+1),\underline{\lambda}},q_{(n+1),\underline{\lambda}})=\exp\left( i(p_{(n+1),\underline{\lambda}} \hat{q}- q_{(n+1),\underline{\lambda}} \hat{p})\right)
\end{equation}
and the conditioned state in the shifted basis is given by
\begin{equation}
\rho_{\ms{c},\underline{\lambda}}^{(n+1)} \rightarrow \hat{D}(p_{(n+1),\underline{\lambda}},q_{(n+1),\underline{\lambda}})
\rho_{\ms{c},\underline{\lambda}}^{(n+1)}\hat{D}(p_{(n+1),\underline{\lambda}},q_{(n+1),\underline{\lambda}})^{\dagger}
\end{equation}

\section{Results}\noindent

The Duffing Hamiltonian (\ref{DuffHam}) has four classical parameters but we will fix the measurement strength so that $\Gamma=0.125$ and we will concentrate on the estimation of the other three parameters: the linear oscillator frequency $\omega$, the coefficient of the nonlinear term $\mu$, and the magnitude of the drive term $g$. The estimated values for these three parameters are denoted by $\tilde{\omega}$, $\tilde{\mu}$, and $\tilde{g}$ respectively. For all of the examples shown below, the actual values for parameter values were set to be $\omega = 1.2$, $\mu = 0.15$, and $g=3.0$. The numerical integration of the SMEs was performed using time steps $\Delta t = 2\pi/500$ so that there were 500 steps per period of the drive term. The individual SMEs for each particle/sample point used a moving basis with 15 harmonic oscillator states, and the composite state was calculated by combining the density matrices from the individual SMEs using (\ref{full_rho}), using a moving basis with 60 harmonic oscillator states.

Figure 2 shows two examples for the estimation of the linear oscillator frequency $\tilde{\omega}$. The examples correspond to the same stochastic record (i.e. the same realization) but with different measurement efficiencies. The blue lines correspond to the case where the measurement is 100\% efficient (with $\eta = 1$). This shows a rapid convergence to the actual value, $\omega= 1.2$, within about 50-100 periods/cycles of the drive term. The 3 sigma errors predicted for the estimate are also shown, together with the resampling events as blue circles. The convergence is fairly rapid and the estimate is relatively stable once converged. The red line on the same figure shows an example where the measurement is inefficient, corresponding to a measurement efficiency of 40\% or $\eta = 0.4$ (chosen to match the estimated efficiency reported in \cite{Web2014}). In this case, the convergence is much slower, indicating that the measurement record contains less information upon which a parameter estimate can be constructed. In this case, the estimated parameter value only stabilizes after around 150-200 cycles of the drive term, and the larger estimated errors indicate this increased uncertainty. In both cases shown, there are slight variations in the estimated values (seen around 200-250 cycles) but these are relatively small and are well within the estimated errors. In addition, where the estimation process takes longer, the number of resampling events (red circles) tends to increase and they often occur later in the process than the corresponding resampling events for efficient measurements, leading to increased computational demands to recalculate the weights after resampling. In addition to the estimates, Figure 2 also shows the purity of the full estimated quantum state for both cases as an inset. For efficient measurements, the conditioned quantum state purifies very rapidly (1-2 periods of the drive term) and remains pure throughout the estimation process. For inefficient measurements, the conditioned quantum state purifies somewhat but then the purity fluctuates between 0.8 and 0.9. The state remains mixed because information about the quantum state is being corrupted by extraneous noise. This is a characteristic of inefficient measurements in quantum systems, and it is not affected by, and does not itself affect, the classical parameter estimation process. 
\begin{figure}[htbp]\label{Fig_2}
	\centering
		\includegraphics[width=1.05\hsize]{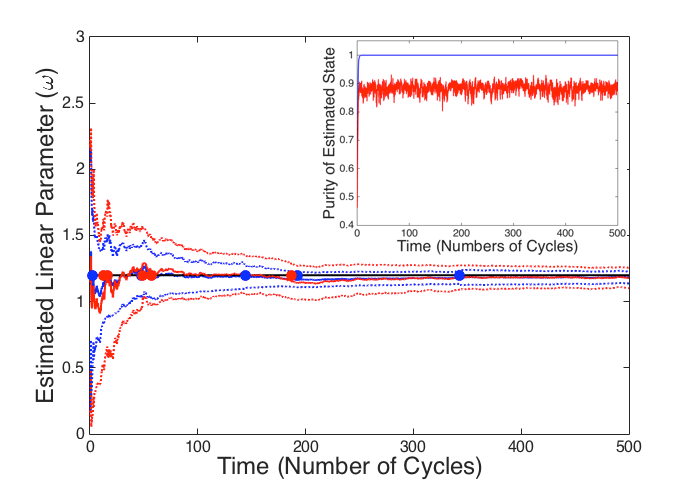}
	\caption{\label{fig:LinearParameter} (Color online) Examples of estimated values for the linear parameter ($\tilde{\omega}$) using SMC sampler with efficient measurements ($\eta = 1.0$, solid blue line) and inefficient measurements ($\eta = 0.4$, solid red line) with an actual linear parameter value $\omega = 1.2$ (solid black line) and 101 sample points (other parameters are given in the text). Three standard deviation errors are indicated in each case with dotted lines, and the resampling points are indicated by circles along the solid black line. The inset figure shows the purity values for the estimated state in each case.}
\end{figure}

Figure 3 shows the evolution of the effective number of particles $N_{eff}$ as a function of time for the examples shown in Figure 2. The resampling events are marked on Figure 2 as large dots, but they are also seen in Figure 3 as large jumps in $N_{eff}$ after the resampling. The data in this figure is useful when optimizing the resampling parameters. It provides information regarding the average number of particles being used. An efficient SMC process would expect to have rapid fluctuations in $N_{eff}$ in the initial phases of the estimation process, with frequent resampling, which would become more gradual drops in $N_{eff}$ as the estimates improve. As time increases, and more measurements are added, the resampling events become less frequent, as is shown in Figures 2 and 3.
\begin{figure}[htbp]\label{Fig_3}
	\centering
		\includegraphics[width=1.05\hsize]{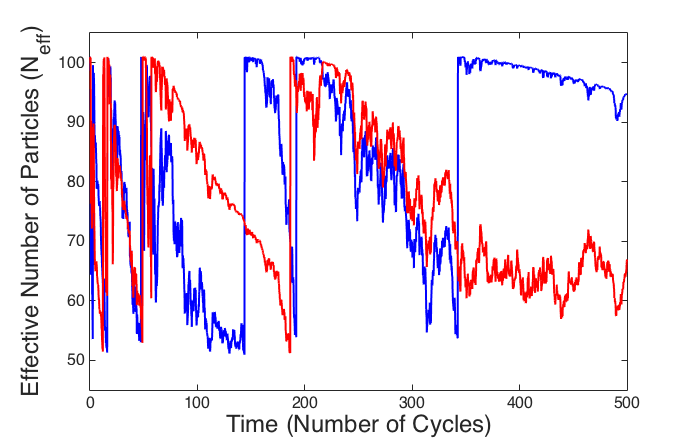}
	\caption{\label{fig:LinearParameter2} (Color online) Examples of the effective number of particles $N_{eff}$ for the estimates shown in Fig.2 for efficient measurements ($\eta = 1.0$, solid blue line) and inefficient measurements ($\eta = 0.4$, solid red line).}
\end{figure}

The estimation of the frequency of the linear oscillator term is relatively straightforward, and this is also found to be the case for the magnitude of the drive term $g$. Estimating the coefficient of the nonlinear term $\mu$ is more challenging however. When the external drive is very small, the Duffing oscillator will appear to be approximately linear and estimating the degree of nonlinearity is problematic. As the amplitude of the drive is increased, the system will explore more of the nonlinear potential and $\mu$ will become easier to estimate. This fact is reflected in the results obtained. For the parameter values selected, the drive term is sufficiently strong to explore the nonlinearity of the potential, but not sufficiently strong so as to require very large numbers of basis states or to make the estimation process easy compared to the other two parameters. An example of the estimation of the nonlinear coefficient is shown in Figure 4, where the convergence to a stable value takes much longer than either example shown in Figure 2, requiring over 500 periods of the drive term to stabilize the estimated value (note the different x-axis compared to Figure 2).
\begin{figure}[htbp]\label{Fig_4}
	\centering
		\includegraphics[width=1.05\hsize]{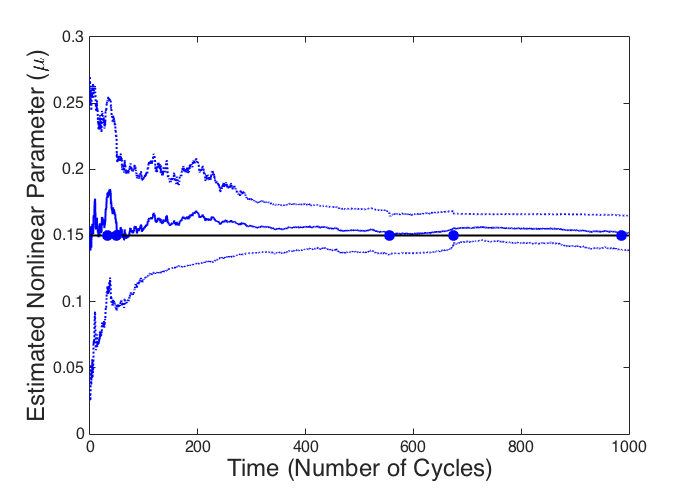}
	\caption{\label{fig:NonLinearParameter} (Color online) An example of estimated values for the nonlinear parameter ($\tilde{\mu}$) using SMC sampler with efficient measurements ($\eta = 1.0$, solid blue line) with an actual nonlinear parameter value  $\mu = 0.15$ (solid black line) and 101 sample points (other parameters are given in the text). Three standard deviation errors are indicated in each case with dotted lines, and the resampling points are indicated by circles along the solid black line.}
\end{figure}
\begin{figure}[htbp]\label{Fig_5}
	\centering
		\includegraphics[width=1.05\hsize]{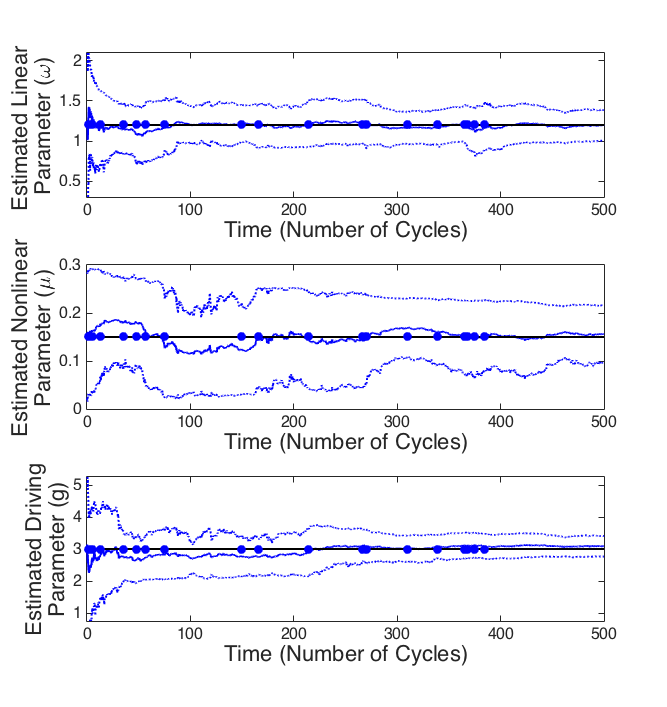}
	\caption{\label{fig:MultiParameter} (Color online) An example of values for all three parameters ($\tilde{\omega}$, $\tilde{\mu}$, and $\tilde{g}$) estimated simultaneously using SMC sampler with efficient measurements ($\eta = 1.0$, solid blue line) and 1001 sample points (other parameters are given in the text). Three standard deviation errors are indicated in each case with dotted lines, and the resampling points are indicated by circles along the solid black line.}
\end{figure}

Each of the examples shown in Figures 2 and 4 show the estimation of one parameter, the other parameters are assumed to be known. The estimation of one parameter is relatively straightforward and a value can be found using a grid-based method (as was the case in \cite{Ral2011} and \cite{Bas2015}). The number of particles required for the estimation of $\tilde{\omega}$ and $\tilde{\mu}$ is around 101 sample points in each of the SMC examples shown above. The number of resampling events is around 4-6 in the cases shown in Figures 2 and 4, and the maximum number of quantum trajectories that would need to be calculated is approximately equivalent to 200-300 trajectories on a fixed grid Hybrid SME. The expected errors for a fixed grid approach are related to the grid spacing, which is related to the initial range over which these grid points are initially distributed. For the cases considered here, with an initial distribution of points for a parameter value $\lambda$ between $0.5\lambda \leq \lambda^{(i)} \leq 1.5\lambda$. Assuming that the actual values of $\lambda$ are uniformly distributed across each interval, the expected error for a fixed grid approach with $N_{grid}$ points would be limited by $\sigma_{\lambda,grid} > (\lambda/N_{grid})/\sqrt{12}\simeq 0.1\%-0.15\%\lambda$. This value is achievable only in the long time limit and the actual error is likely to be significantly larger than this. In the examples given above, the SMC sampler produces parameter estimates with errors approaching this limit within a few hundred cycles. There is therefore a small but potentially significant benefit in using the SMC sampler method for one parameter estimation.

Moving from single to multiple parameter estimation presents a serious problem for grid-based methods. The number of points required scales exponentially in the number of dimensions to achieve the same accuracy. The error from a grid-based approach results from approximating an integral of functions in $D$ dimensions, where the error is $O((N_{grid})^{(-1/D)})$. The error for an SMC sampler comes from approximating the integral directly (using Monte-Carlo integration) and therefore is $O(1/N_{grid})$ whatever value $D$ takes \cite{Pre2007}. (See reference [40] for proofs for the convergence of SMC and particle filter based methods). So, for $D=1$, the two approaches offer similar scaling of error with $N_{grid}$, in higher dimensions, an SMC sampler will asymptotically outperform a grid-based method as $N_{grid}$ tends to infinity. Of course, differences in constants of proportionality mean that a computational benefit from using the SMC sampler in a small number of dimensions (number of parameters) is not guaranteed. Estimating all three parameters in our example, at a level of accuracy equivalent to the one parameter examples above, would require around ten million grid points, $(300)^3 = 9\times 10^6$. With SMC methods, this number is dramatically reduced. 

Figure 5 shows an example of the simultaneous estimation of all three parameters using 1001 sample points. The values for $\tilde{\omega}$ and $\tilde{g}$ still converge rapidly whilst $\tilde{\mu}$ takes longer to establish a stable estimate. When comparing this with a grid based method, we note that the number of trajectories is larger than for the single parameter case and the number of resampling events is also increased, approximately 20 in the case shown in Figure 5. This is equivalent to a run-time for approximately 10,000 trajectories on a fixed grid. Using the same assumptions as before, this would give errors limited by $\sigma_{\underline{\lambda},grid} > (\lambda/\sqrt[3]{N_{grid}})/\sqrt{12}\simeq 1.5\% \lambda$. The errors found using the SMC sampler described above are nearly an order of magnitude smaller than this limit for one of the three parameters ($\omega$) and comparable for the remaining two parameters ($\mu$ and $g$). There is an additional benefit, in that the sample points not only provide estimates of the parameter values, they also provide information regarding the correlations between the different parameters. For the example shown in Figure 5, the mean vector and the estimated covariance matrix ($S$) are given by
$$
\left(\begin{array}{c} \tilde{\omega} \\ \tilde{\mu} \\ \tilde{g} \end{array} \right) = \left(\begin{array}{c} 1.1981 \\ 0.1557 \\ 3.0874 \end{array}  \right) 
$$
$$ \small{
S = \left( \begin{array}{ccc} 4.0308 \times 10^{-3} & -1.6449 \times 10^{-6}  & 9.2270 \times 10^{-6} \\ -1.6449 \times 10^{-6} & 3.9795 \times 10^{-4}  & -2.8254  \times 10^{-6}  \\ 9.2270 \times 10^{-6} & -2.8254  \times 10^{-6} & 1.1586 \times 10^{-2} \end{array}\right)}
$$
Note also that in Figure 5, the standard deviation of the linear parameter ($\tilde{\omega}$) is larger than in Figure 2 and the convergence is slower than for the single parameter case. This is partly due to the larger uncertainty generally in the three unknown parameters, and in part due to the slower convergence of the nonlinear parameter ($\tilde{\mu}$). The coupling between the parameters, shown by the non-negligible correlations shown in the covariance matrix, means that uncertainty in the nonlinear parameter increases the standard deviation of the other two parameters.

The use of an SMC sampler to estimate the Hamiltonian parameter values directly from the quantum trajectories is more efficient than an equivalent grid-based method but it still presents a computational challenge. Solving a single SME can be simplified using a stochastic integration method designed specifically for SMEs, like Rouchon's method \cite{Ami2011,Rou2015}, and using efficient numerical tools, like moving basis states \cite{Sch1995,Bru1996}. However, solving many simultaneous SMEs to determine the evolution of the particle weights still requires significant computational resources. The number of combinations of parameter values explored using the SMC sampler is significantly less than that required by a conventional grid-based method, but each sample point explored requires the full trajectory to calculated, or recalculated after resampling. The number of SMEs required to be calculated can be said to be {\em relatively} small but it is still not a trivial exercise. In their favor, SMC methods are amenable to parallelization \cite{Gre2017}, since the evolution of SME and the recalculation of each trajectory after resampling are largely independent processes and can be distributed simply across a number of processors. However, at present, it is more likely that this type of technique is more likely to be used for post-processing experimental data rather than as part of an on-line closed-loop control system.  

\section{Conclusions}\noindent

Continuous quantum measurements, and their associated stochastic master equations (SMEs), provide a means to monitor the dynamical evolution of a quantum system and to provide an estimate of the underlying quantum state. In addition, the quantum trajectories resulting from the integration of stochastic master equations contain useful information about the parameters that govern the evolution of the system. Hybrid stochastic master equations provide a means to extract the information regarding these classical parameters. Hybrid SMEs involve running many parallel SMEs, each one having a different value for the parameter (or parameters). The classical probabilities attached to the individual SMEs and the associated parameter values can then be found by integrating a Kushner-Stratonovich equation. This classical estimation process is numerically costly, and is even more so when estimates are required for multiple parameters. This paper has demonstrated how such estimates can be found using a technique taken from classical state estimation and nonlinear filtering, a Sequential Monte Carlo (SMC) sampler. The SMC sampler used in this paper has been demonstrated to allow the simultaneous estimation of three Hamiltonian parameters, together with their statistical correlation and the associated quantum trajectories, in a computationally tractable form, with a relatively small number of candidate parameter values and parallel SMEs.  

Even with such methods, the computational task in solving the Hybrid SME is formidable, and is currently beyond the point where it could be used as part of a closed-loop quantum control system. At present, the strength of such techniques is in the ability to post-process experimental measurement data to verify the quantum states used in an experiment but also to provide an independent, in-situ means to check the parameters that govern their evolution.  

\vspace{5mm}

\textit{Acknowledgments:} JFR would like to thank the US Army Research Laboratories (contract no. W911NF-16-2-0067). JFR would also like to thank Hendrik Ulbricht and Peter Barker for helpful and informative discussions.

\bibliographystyle{apsrev}

\begin{thebibliography}{99}

\bibitem{Bel1999} V. P. Belavkin {\em Reports on Mathematical Physics} {\bf 45}, 353 (1999), and references contained therein.
\bibitem{Wis2010} H. M. Wiseman, G. J. Milburn, `Quantum Measurement and Control' (Cambridge University Press, Cambridge, 2010).
\bibitem{Jac2014} K. Jacobs `Quantum Measurement Theory and Its Applications' (Cambridge University Press, Cambridge, 2014). 
\bibitem{Mur2013} K. W. Murch, S. J. Weber, C. Macklin, I. Siddiqi, {\em Nature} {\bf 502}, 211 (2013).
\bibitem{Web2014} S. J. Weber, A. Chantasri, J. Dressel, A. N. Jordan, K. W. Murch, I. Siddiqi, {\em Nature} {\bf 511}, 570 (2014).
\bibitem{Six2015} P. Six, P. Campagne-Ibarcq, L. Bretheau, B. Huard, P. Rouchon, {\em 54th IEEE Conference on
Decision and Control Conference (CDC)} (2015).
\bibitem{Cam2016} P. Campagne-Ibarcq, P. Six, L. Bretheau, A. Sarlette, M. Mirrahimi, P. Rouchon, B. Huard, {\em Physical Review X} {\bf 6}, 011002 (2016).
\bibitem{Smi2002} P. Smith, J. E. Reiner, L. A. Orozco, S. Kuhr, H. M. Wiseman, {\em Physical Review Letters} {\bf 89},  133601 (2002). 
\bibitem{Bra2012} S. Brakhane, W. Alt,  T. Kampschulte,  M. Martinez-Dorantes, R. Reimann, S. Yoon,  A. Widera,  D. Meschede, {\em Physical Review Letters} {\bf 109}, 173601 (2012). 
\bibitem{Kub2009} Kuban A. Kubanek, M. Koch, C. Sames, A. Ourjoumtsev, P. W. H. Pinkse, K. Murr, G. Rempe, {\em Nature} {\bf 462}, 898, (2009). 
\bibitem{Wil2015} D.J. Wilson, V. Sudhir, N. Piro, R. Schilling, A. H. Ghadimi, and T.J. Kippenberg, {\em Nature}, {\bf 524}, 325 (2015). 
\bibitem{Sud2017} V. Sudhir, D.J. Wilson, R. Schilling, H. Sch{\"u}tz, S. A. Fedorov, A. H.  Ghadimi, A. Nunnenkamp, and T.J. Kippenberg, {\em Physical Review X} {\bf 7}, 011001 (2017). 
\bibitem{Vij2012} R. Vijay, C. Macklin, D.H. Slichter, S.J. Weber,  K.W. Murch, R. Naik, A. N. Korotkov, I. Siddiqi, {\em Nature} {\bf 490}, 77 (2012). 
\bibitem{Ris2012} D. Rist{\`e},C.C. Bultink, K.W. Lehnert, and L. DiCarlo, {\em Physical Review Letters} {\bf 109}, 240502 (2012). 
\bibitem{Ral2011} J. F. Ralph, K. Jacobs, C. D. Hill, {\em Physical Review A} {\bf 84}, 052119 (2011).
\bibitem{Gre2017} P. L. Green, S Maskell, {\em Mechanical Systems and Signal Processing}, {\bf 93}, 379Ð396 (2017).
\bibitem{Chu1997} I. L. Chuang, M. A. Nielsen, {\em Journal Modern Optics} {\bf 44}, 2455 (1997).
\bibitem{Gam2001} J. Gambetta, H. M. Wiseman, {\em Physical Review A} {\bf 64}, 042105 (2001).
\bibitem{Ver2001} F. Verstraete, A. C. Doherty, H. Mabuchi, {\em Physical Review A} {\bf 64}, 032111 (2001).
\bibitem{Sto2004}  J. K. Stockton, J. M. Geremia, A. C. Doherty, H. Mabuchi, {\em Physical Review A} {\bf 69}, 032109 (2004).
\bibitem{Tsa2009a} M. Tsang, {\em Physcial Review Letters} {\bf 102}, 250403 (2009).        
\bibitem{Tsa2009b} M. Tsang, {\em Physical Review A} {\bf 80}, 033840 (2009).          
\bibitem{Tsa2010} M. Tsang, {\em Physical Review A} {\bf 81}, 013824 (2010).     
\bibitem{Tsa2011} M. Tsang, H.M. Wiseman, C.M. Caves, {\em Physical Review Letters} {\bf 106}, 90401 (2011). 
\bibitem{Neg2013} A Negretti, K. {M\/{o}lmer}, {\em New Journal Physics} {\bf 15}, 125002 (2013). 
\bibitem{Ber2015} D.W. Berry, M. Tsang, M.J.W. Hall, H.M. Wiseman, {\em Physical Review X} {\bf 5}, 031018 (2015). 
\bibitem{Bas2015} H. Bassa,  S.K. Goyal, S.K. Choudhary, H. Uys, L. {Di\'{o}}si, T. Konrad, {\em Physical Review A} {\bf 92}, 032102 (2015). 
\bibitem{Cor2017} L. Cortez, A. Chantasri, L.P. Garc\'{\i}a-Pintos, J. Dressel, A.N. Jordan, {\em Physical Review A} {\bf95}, 012314 (2017). 
\bibitem{LIGO2016} B.P. Abbott, R. Abbott, T.D. Abbott, \textit{et al.}, {\em Physical Review Letters} {\bf116}, 061102 (2016). 
\bibitem{Yan2017} F. Yang, A.J. Koll{\'{a}}r, S.F. Taylor, R.W. Turner, B.L. Lev, {\em Physical Review Applied} {\bf 7}, 034026 (2017). 
\bibitem{Gor1993} N. Gordon, D. Salmond, A. F. Smith, {\em IEE Proceedings F Radar Signal Processing}, {\bf 140}, 107Ð113 (1993).
\bibitem{Bla1986} S. Blackman, `Multiple Target Tracking with Radar Applications' (Artech House, 1986)
\bibitem{Bar2001} Y. Bar-Shalom, X.R. Li, T. Kirubarajan, `Estimation with Applications to Tracking and Navigation' (Wiley \& Sons, 2001).
\bibitem{Ral2010} J. F. Ralph, `Target Tracking' in `Encyclopedia of Aerospace Engineering', Vol.5, Ch 251, eds. R. Blockley, W. Shyy (Wiley \& Sons, 2010).
\bibitem{Wis2005} H. M. Wiseman, A. Doherty, {\em Physical Review Letters} {\bf 94}, 070405 (2005).
\bibitem{Dou2001} A. Doucet, N. De Freitas, and N. Gordon, Eds., `Sequential Monte Carlo Methods in Practice' (Springer, New York, 2001).
\bibitem{Cap2007} O. Capp\'{e}, S. J. Godsill, E. Moulines, {\em Proceedings of the IEEE} {\bf 95}, 899 (2007).
\bibitem{Aru2002} M. Arulampalam, S. Maskell, N. Gordon, T. Clapp, {\em IEEE Transactions on Signal Processing} {\bf 50}, 241-254, (2002).
\bibitem{Pre2007} W. H. Press, S. A. Teukolsky, W. T. Vetterling, B. P. Flannery, ``Section 7.9.1 Importance Sampling" in ``Numerical Recipes: The Art of Scientific Computing (3rd ed.)" (Cambridge University Press, New York, 2007).
\bibitem{Cri2002} D. Crisan, A. Doucet. {\em IEEE Transactions on signal processing} {\bf 50},  736-746 (2002).
\bibitem{Hes1995} T. Hesterberg, {\em Technometrics}, {\bf 37}, 185-194 (1995).
\bibitem{Sch1995} R. Schack, T. A. Brun, I. C. Percival. {\em Journal of Physics A: Mathematical and General} {\bf 28}, 5401 (1995).
\bibitem{Bru1996} T. A. Brun, I. C. Percival, R. Schack, {\em Journal of Physics. A: Mathematical and General} {\bf 29} 2077 (1996).
\bibitem{Bru1997} T. A. Brun, N. Gisin, P. F. O'Mahony, M. Rigo, {\em Physics Letters A} {\bf 229} 267-272 (1997).
\bibitem{Hab1998} S. Habib, K. Shizume, W. H. Zurek, {\em Physical Review Letters} {\bf 80}, 4361 (1998).
\bibitem{Bha2000} T. Bhattacharya, S. Habib, K. Jacobs, {\em Physical Review Letters} {\bf 85} 4852 (2000).
\bibitem{Sco2001} A. J. Scott, G. J. Milburn, {\em Physical Review A} {\bf 63}, 042101 (2001).
\bibitem{Bha2003} T. Bhattacharya, S. Habib, K. Jacobs, {\em Physical Review A} {\bf 7} 042103 (2003).
\bibitem{Eve2005} M. J. Everitt, T. D. Clark, P. B. Stiffell, J. F. Ralph, A. R. Bulsara, C. J. Harland. {\em New Journal of Physics} {\bf 7} 64 (2005).
\bibitem{Eas2016} J. K. Eastman, J. J. Hope, A. R. R. Carvalho. {\em Emergence of chaos controlled by quantum noise}, arXiv:1604.03494 (2016).
\bibitem{Pok2016} B. Pokharel, P. Duggins, M. Misplon, W. Lynn, K. Hallman, D. Anderson, A. Kapulkin, A. K. Pattanayak, {\em Dynamical complexity in the quantum to classical transition}, arXiv:1604.02743 (2016).
\bibitem{Ral2017} J. F. Ralph, K. Jacobs, M. J. Everitt, {\em Physical Review A} {\bf 95}, 012135 (2017).
\bibitem{Gie2013} J. Gieseler, L. Novotny, R. Quidant, {\em Nature Physics}, {\bf 9}, 806 (2013).
\bibitem{Gie2015} J. Gieseler, L. Novotny, C. Moritz, C. Dellago, {\em New Journal of Physics}, {\bf 17}, 045011 (2015).
\bibitem{Aik2001} M. Aikele, K. Bauer, W. Ficker, F. Naubauer, U. Prechtel, J. Schalk, H. Seidel, {\em Sensors and Actuators A} {\bf 90}, 161-167 (2001).
\bibitem{Mes2009} R. M. C. Mestrom, R. H. B. Fey, H. Nijmeijer, {\em IEEE/AMSE Transactions on Mechatronics} {\bf 14}, 423-433 (2009).
\bibitem{Agr2013} D. K. Agrawal, J. Woodhouse, A. A. Seshia, {\em IEEE Trans. on Ultrasonics, Ferroelectrics and Frequency Control}, {\bf 60}, 1646-1659 (2013).
\bibitem{Ami2011} H. Amini, M. Mirrahimi, P. Rouchon, {\em in Proc. 50th IEEE Conf. on Decision and Control},  pp. 6242-6247 (2011).
\bibitem{Rou2015} P. Rouchon, J. F. Ralph, {\em Physical Review A} {\bf 91}, 012118 (2015).
\bibitem{Mil1995} G.N. Milstein, `Numerical Integration of Stochastic Differential Equations' (Springer, Berlin, 1995).
\bibitem{Ral2016} J. F. Ralph, K. Jacobs, J. Coleman, {\em Physical Review A} {\bf 94}, 032108 (2016).

\end{thebibliography}

\end{document}